\documentclass[10pt]{article}

\usepackage{amsmath}
\usepackage{amssymb}

\usepackage{graphicx}

\usepackage{cite}

\usepackage{color} 

\usepackage[labelfont=bf,labelsep=period,justification=raggedright]{caption}

\makeatletter
\renewcommand{\@biblabel}[1]{\quad#1.}
\makeatother

\date{}

\begin{document}

% Title must be 150 characters or less
\begin{flushleft}
{\Large
\textbf{Statistical Basis for Predicting Technological Progress}
}
% Insert Author names, affiliations and corresponding author email.
\\
B\'ela Nagy$^{1,\ast}$, 
J. Doyne Farmer$^{1}$, 
Quan M. Bui$^{1,2}$
Jessika E Trancik$^{1,3,\ast}$
\\
\bf{1} Santa Fe Institute, 1399 Hyde Park Road, Santa Fe, NM 87501, USA
\\
\bf{2} St. John's College,1160 Camino Cruz Blanca, Santa Fe, NM, 87505, USA
\\
\bf{3} Engineering Systems Division, Massachusetts Institute of Technology, Cambridge, MA 02139, USA
\\
$\ast$ E-mail: Corresponding bn@santafe.edu,  trancik@mit.edu
\end{flushleft}

% Please keep the abstract between 250 and 300 words
\section*{Abstract}
Forecasting technological progress is of great interest to engineers, policy makers, and private investors. Several models have been proposed for predicting technological improvement, but how well do these models perform? An early hypothesis made by Theodore Wright in 1936 is that cost decreases as a power law of cumulative production. An alternative hypothesis is Moore's law,  which can be generalized to say that technologies improve exponentially with time.  Other alternatives were proposed by Goddard, Sinclair et al., and Nordhaus.  These hypotheses have not previously been rigorously tested. Using a new database on the cost and production of 62 different technologies, which is the most expansive of its kind, we test the ability of six different postulated laws to predict future costs.  Our approach involves hindcasting and developing a statistical model to rank the performance of the postulated laws. Wright's law produces the best forecasts, but Moore's law is not far behind. We discover a previously unobserved regularity that production tends to increase exponentially. A combination of an exponential decrease in cost and an exponential increase in production would make Moore's law and Wright's law indistinguishable, as originally pointed out by Sahal. We show for the first time that these regularities are observed in data to such a degree that the performance of these two laws is nearly tied. Our results show that technological progress is forecastable, with the square root of the logarithmic error growing linearly with the forecasting horizon at a typical rate of 2.5\% per year. These results have implications for theories of technological change, and assessments of candidate technologies and policies for climate change mitigation.

% Please keep the Author Summary between 150 and 200 words
% Use first person. PLoS ONE authors please skip this step. 
% Author Summary not valid for PLoS ONE submissions.   
%\section*{Author Summary}

\section*{Introduction}
Innovation is by definition new and unexpected, and might therefore seem inherently unpredictable. But if there is a degree of predictability in technological innovation, understanding it could have profound implications. Such knowledge could result in better theories of economic growth, and enable more effective strategies for engineering design, public policy design, and private investment. In the area of climate change mitigation, the estimated cost of achieving a given greenhouse gas concentration stabilization target is highly sensitive to assumptions about future technological progress \cite{Gillingham08}. 

There are many hypotheses about technological progress, but are they any good? Which, if any, hypothesis provides good forecasts? In this paper, we present the first statistically rigorous comparison of competing proposals.

When we think about progress in technologies, the first product that comes to mind for many is a computer, or more generally, information technologies.   The following quote by Bill Gates captures a commonly held view:  ``Exponential improvement -- that is rare -- we've all been spoiled and deeply confused by the IT model''\cite{Fried}. But as we demonstrate here, information technologies are not special in terms of the functional form that describes their improvement over time. Information technologies show rapid rates of improvement, but many technologies show exponential improvement.  In fact, all the technologies we study here behave roughly similarly:  Information technologies closely follow patterns of improvement originally postulated by Wright for airplanes \cite{Wright36,Argote90,Dutton84,IEA00,McDonald01,Alberth08}, and technologies such as beer production or offshore gas pipelines follow Moore's law \cite{Moore65,Moore75}, but with a slower rate of improvement \cite{Alberth08,Kurzweil2,Koh06,Nordhaus07,Koh08,Amaya08}.

It is not possible to quantify the performance of a technology with a single number \cite{Dosi10}.  A computer, for example, is characterized by speed, storage capacity, size and cost, as well as other intangible characteristics such as aesthetics.  One automobile may be faster while another is less expensive.  For this study we focus on one common measure of performance: the inflation-adjusted cost of one ``unit".  This metric is suitable in that it can be used to describe many different technologies. However, the nature of a unit may change over time. For example, a transistor in a modern integrated circuit today may have quite different performance characteristics than its discrete counterpart in the past. Furthermore, the degree to which cost is emphasized over other performance measures may change with time \cite{Abernathy74}. We nonetheless use the changes in the unit cost as our measure of progress, in order to compare competing models using a sizable dataset. The crudeness of this approach only increases the difficulty of forecasting and makes it particularly surprising that we nonetheless observe common trends.

% You may title this section "Methods" or "Models". 
% "Models" is not a valid title for PLoS ONE authors. However, PLoS ONE
% authors may use "Analysis" 
\section*{Analysis}
We test six different hypotheses that have appeared in the literature \cite{Wright36,Moore65,Goddard82,Sinclair00,Nordhaus09}, corresponding to the following six functional forms:
\begin{eqnarray}
\label{laws}\mbox{Moore}~~\log y_t & =  & a t + b + n(t)\\
\nonumber \mbox{Wright}~~\log y_t & = & a \log x_t + b + n(t)\\
\nonumber \mbox{lagged Wright}~~\log y_t & = & a \log (x_t - q_t) + b + n(t)\\
\nonumber \mbox{Goddard}~~\log y_t & = & a \log q_t + b + n(t)\\
\nonumber \mbox{SKC}~~\log y_t & = & a \log q_t + c \log (x_t - q_t) + b + n(t)\\
\nonumber \mbox{Nordhaus}~~\log y_t & =  & a t + c \log x_t + b + n(t)
\end{eqnarray}
The dependent variable $y_t$ is the unit cost of the technology measured in inflation adjusted dollars. The independent variables are the time $t$ (measured in years), the annual production $q_t$, and the cumulative production $x_t = \sum_{i = 1}^t q_i$.  The noise term $n(t)$, the constants $a$, $b$ and $c$ and the predictor variables differ for each hypothesis.  

{\it Moore's law} here refers to the generalized statement that the cost $y$ of a given technology decreases exponentially with time, i.e. 
\begin{equation}
y_t = B \exp (-mt),
\label{Moore}
\end{equation}
where $m > 0$ and $B > 0$ are constants \cite{Moore65,Koh06}.  (We assume throughout that $t > 0$, and we have renamed $a = -m$ and $b = \log B$ in Eq.~(\ref{laws})).   Moore's law postulates that technological progress is inexorable, i.e. it depends on time rather than controllable factors such as research and development.

{\it Wright's law}, in contrast, postulates that cost decreases at a rate that depends on cumulative production, i.e.
\begin{equation}
y_t = B x_t^{-w},
\label{Wright}
\end{equation}
where $w > 0$ and $B  > 0$ are constants, and we have renamed $a = -w$ and $b = \log B$ in Eq.~(\ref{laws}).   Wright's law is often interpreted to imply ``learning by doing" \cite{Dutton84,Thompson10}.  The basic idea is that cumulative production is a proxy for level of effort, so that the more we make the more we learn, and knowledge accumulates without loss.

Another hypothesis is due to Goddard \cite{Goddard82}, who argues that progress is driven purely by economies of scale, and postulates that: 

\begin{equation}
y_t = B q_t^{-s},
\label{Goddard}
\end{equation}
where $s > 0$ and $B  > 0$ are constants, and we have renamed $a = -s$ and $b = \log B$ in Eq.~(\ref{laws}).  

We also consider the three multi-variable hypotheses in Eq.~(\ref{laws}):  Nordhaus \cite{Nordhaus09} combines Wright's law and Moore's law\footnote{Note that the conclusions presented do not work against Nordhaus' point about the difficulty in separating learning from exogenous sources of change \cite{Nordhaus09}.}, and Sinclair, Klepper, and Cohen (SKC) \cite{Sinclair00} combine Wright's law and Goddard's law.  For completeness we also test Wright's law lagged by one year.  Note that these methods forecast different things:  Moore's law forecasts the cost at a given time, Wright's law at a given cumulative production, and Goddard's law at a given annual production.

We test these hypotheses on historical data consisting of 62 different technologies that can be broadly grouped into four categories: Chemical, Hardware, Energy, and Other\footnote{
All data can be found in the online Performance Curve Database at pcdb.santafe.edu.}.  
The data are sampled at annual intervals with timespans ranging from 10 to 39 years. The choice of these particular technologies was driven by availability -- we included all of the data that we could find to assemble the largest database of its kind.  For a detailed description see the Supporting Information. 

To compare the performance of each hypothesis we use hindcasting, which is a form of cross-validation.  We pretend to be at time $i$ and make a forecast $\hat{y}_j^{(f,d,i)}$ for time $j$ using hypothesis (functional form) $f$ and data set $d$, where $j > i$.  The parameters for each functional form are fit using ordinary least squares based on all data prior to time $i$, and forecasts are made based on the resulting regression\footnote{
An alternative is to adjust the intercepts to match the last point, which produces better short term forecasts.  For example, for Moore's law this corresponds to using a log random walk of the form $\log y_{t+1} = \log y_t - \mu + n(t)$, where $n(t)$ is an IID noise term.  We have not done this here to be consistent with the way these hypotheses have been historically presented.  The method we have used here also results in more stable errors.  Our purpose here is not to propose an optimal forecasting method, but rather to compare existing hypotheses.}.
We score the quality of forecasts based on the logarithmic forecasting error 

\begin{equation}
e_{fdij} = \log y_j^{(d)} - \log \hat{y}_j^{(f,d,i)}
\label{logerror}
\end{equation}
The quality of forecasts is examined for all datasets and all hypotheses (and visualized as a three-dimensional error mountain, as shown in the Supporting Information). For Wright's law an illustration of the growth of forecasting errors as a function of the forecasting horizon is given in Fig. ~\ref{errorVsTime}.

Developing a statistical model to compare the competing hypotheses is complicated by the fact that errors observed at longer horizons tend to be larger than those at shorter horizons, and errors are correlated across time and across functional forms.  After comparing many different possibilities (as discussed in detail in the Supporting Information), we settled on the following approach. Based on a search of the family of power transformations, which is known for its ability to accommodate a range of variance structures, we take as a response the square root transformation of the logarithmic error. This response was chosen to maximize likelihood when modeled as a linear function of the hindcasting horizon $\:=\:$ target $\:-\:$ origin $\:=\: j-i$, using a linear mixed effects model. 

Specifically, we use the following functional form to model the response. 
\begin{equation}
r_{fdij} \equiv | e_{fdij} |^{0.5} = \alpha_f + a_d + (\beta_f + b_d) (j - i) + \epsilon_{fdij},
\label{errorModel}
\end{equation}
where $r_{fdij}$ is the expected root error.  The parameters $\alpha_f$ and $\beta_f$ depend on the functional form and are called {\it fixed effects} because they are the same for all datasets.  $\alpha_f$ is the intercept and $\beta_f $ is the slope parameter. 

The parameters $a_d$ and $b_d$ depend on the dataset, and are called {\it random effects} because they are not fitted independently, but are instead treated as dataset-specific random fluctuations from the pooled data. The quantities $a_d$ and $b_d$ are additive adjustments to the average intercept and slope parameters $\alpha_f$ and $\beta_f$, respectively, to take into account the peculiarities of each dataset $d$.

In order to avoid adding 62 $a_d$ parameters plus 62 $b_d$ parameters, we treated the $\begin{pmatrix} a_d \\ b_d \end{pmatrix}$ pair as a two-dimensional random vector having a bivariate normal distribution with mean $\begin{pmatrix} 0 \\ 0 \end{pmatrix}$ and variance-covariance matrix $\begin{pmatrix} \psi_a^2 & \psi_{ab} \\ \psi_{ab} & \psi_b^2 \end{pmatrix}$. This approach dramatically reduces the number of parameters. We parameterize the dataset-specific adjustments as random deviations from the average $\begin{pmatrix} \alpha_f  \\ \beta_f  \end{pmatrix}$ at a cost of only 3 additional parameters instead of 2 $\times$ 62 $=$ 124. This parsimonious approach makes maximum likelihood estimation possible by keeping the number of parameters in check.

Finally, we add an $\varepsilon_{fdij}$ random field term to take into account the deviations from the trend. This is assumed to be a Gaussian stochastic process independent of the $\begin{pmatrix} a_d \\ b_d \end{pmatrix}$ random vector, having mean $0$, and given $a_d$ and $b_d$, having variance equal to a positive $\sigma^2$ times the fitted values:
\begin{equation} \mbox{Var}\left( \left. \varepsilon_{fdij} \right| a_d, b_d \right) \:=\: \sigma^2\: \mbox{E}\left( \left. r_{fdij} \right| a_d, b_d \right) \label{Var}\end{equation}
We also define an exponential correlation structure within each error mountain (corresponding to each combination of dataset and hypothesis, see the Supporting Information), as a function of the differences of the two time coordinates with a positive range parameter $\rho$ and another small positive nugget parameter $\eta$ quantifying the extent of these correlations:\\
\begin{equation} \mbox{Corr}(\varepsilon_{fdij},\: \varepsilon_{f'd'i'j'}) \:=\: \delta_{ff'} \delta_{dd'} (1-\eta) \exp{\left\{-(|i-i'|+|j-j'|)/\rho \right\}}, \label{Corr}\end{equation}
where the two Kronecker $\delta$ functions ensure that each error mountain is treated as a separate entity.

Equations (\ref{Var}) and (\ref{Corr}) were chosen to deal with the observed heteroscedasticity (increasing variance with increasing logarithmic forecasting error) and the serial correlations along the time coordinates $i$ (hindcasting origin) and $j$ (hindcasting target). Based on the likelihood, an exponential correlation function provided the best fit. Note that instead of a Euclidean distance (root sum of the squares of differences), the Manhattan measure was used (the sum of the absolute differences), because it provided a better fit in terms of the likelihood.
 
Using this statistical model, we compared five different hypotheses. (We removed the Nordhaus model from the sample because of poor forecasting performance. This model gave good in-sample fits but generated large and inconsistent errors when predicting out-of-sample, a signature of over-fitting.) Rather than the $62 \times 5 \times 2 = 620$ parameters needed to fit each of the 62 datasets separately for each of the five functional forms, there are only $16$ free parameters:  $5 \times 2$ = 10 parameters $\alpha_f$ and $\beta_f$,  three parameters for the covariance matrix of the bivariate random vector $(a_d, b_d)$, and three parameters for the variance and autocorrelation of the residuals $\epsilon_{fdij}$. 

% Results and Discussion can be combined.
\section*{Results and Discussion}
\begin{figure}
\vspace{2em}
\centering
\hspace{-2em}
\includegraphics[scale=0.45]{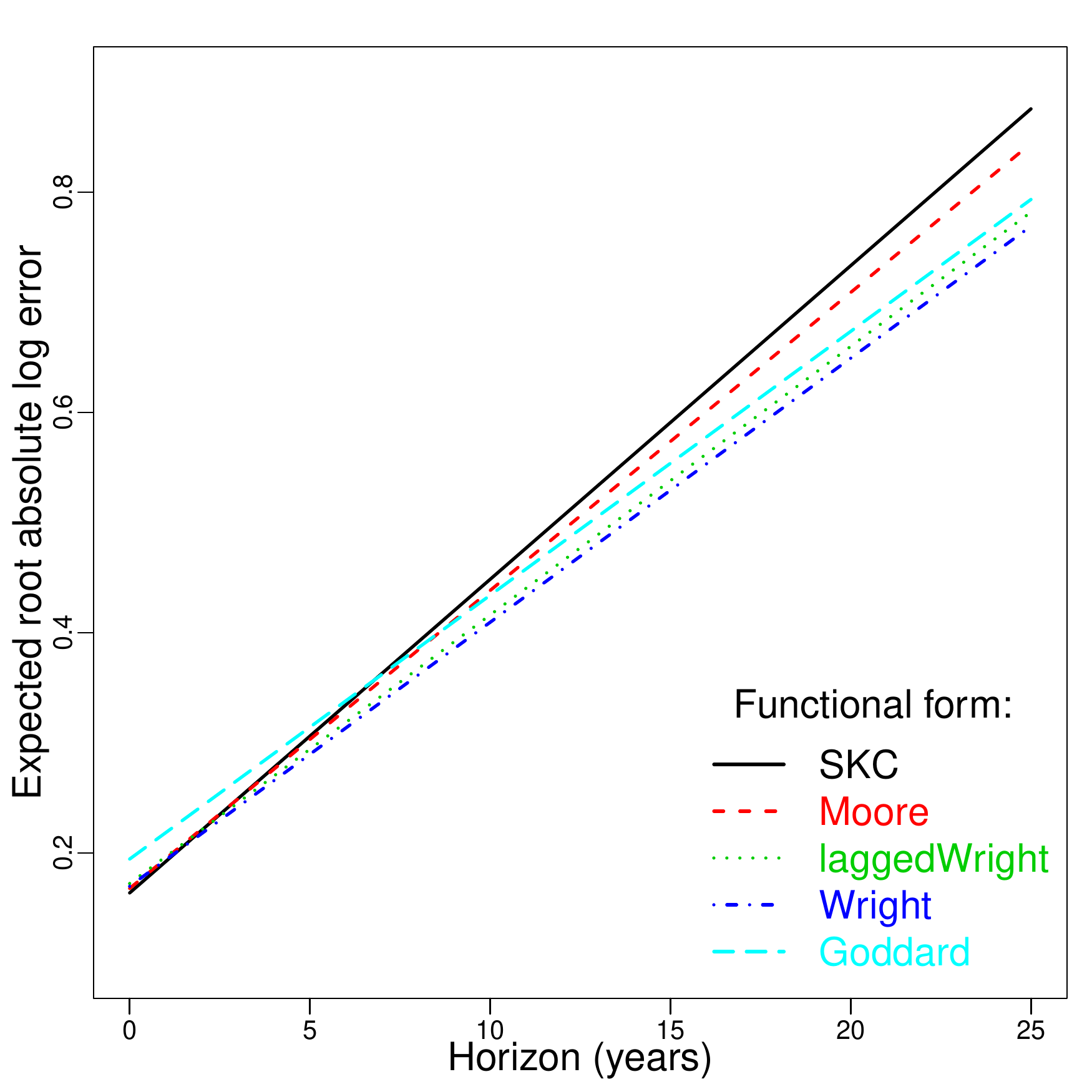} 
\caption{{\bf An illustration of the growth of errors of each hypothesized law vs. time.}  The plot shows the predicted root absolute log error $r_{fij}$ vs. forecasting horizon $(j - i)$ using each of the functional forms (see Eq.~(\ref{errorModel})).  The performance of the five hypotheses shown is fairly similar, though Goddard is worse at short horizons and SKC and Moore are worse at long horizons.}
\label{fits}
\end{figure}

We fit the error model to the $37,745$ different $r_{fdij}$ data points using the method of maximum likelihood. In Fig.~\ref{fits} we plot the expected root error $r_{fij} = \alpha_f + \beta_f (j - i)$ for the five hypotheses as a function of the hindcasting horizon. While there are differences in the performance of these five hypotheses, they are not dramatic. The intercept is tightly clustered in a range $0.16 < \alpha_f < 0.19$ and the slope $0.024 < \beta_f < 0.028$.  Thus all the hypotheses show a large initial error, followed by a growth in the root error of roughly $2.5\%$ per year\footnote{This is a central tendency for the pooled data.}.

The error model allows us to compare each hypothesis pairwise to determine whether it is possible to reject one in favor of another at statistically significant levels.  The comparisons are based on the intercept and slope of the error model of Eq.~(\ref{errorModel}). The parameter estimates are listed in Tables S1 and S3 and the corresponding $p$-values in Tables S2 and S4.  For example, at the 5\% level, the intercept of Goddard is significantly higher than any of the others and the slope of SKC is significantly greater than that of Wright, lagged Wright and Goddard. With respect to slope, Moore is at the boundary of being rejected in favor of Wright.  Fig.~\ref{fits} makes the basic pattern clear:  Goddard does a poorer job of forecasting at short times, whereas SKC and to a lesser extent Moore do a poorer job at long times.  

We thus have the surprising result that most of the methods are quite similar in their performance.  Although the difference is not large, the fact that we can eliminate Goddard for short term forecasts indicates that there is information in the cumulative production not contained in the annual production, and suggests that there is a learning effect in addition to economies of scale. But the fact that Goddard is not that much worse indicates that much of the predictability comes from annual production, suggesting that economies of scale are important. (In our database technologies rarely decrease significantly in annual production; examples of this would provide a better test of Goddard's theory.)  We believe the SKC model performs worse at long times because it has an extra parameter, making it prone to overfitting.

\begin{figure}
\centering{\includegraphics[scale=0.5]{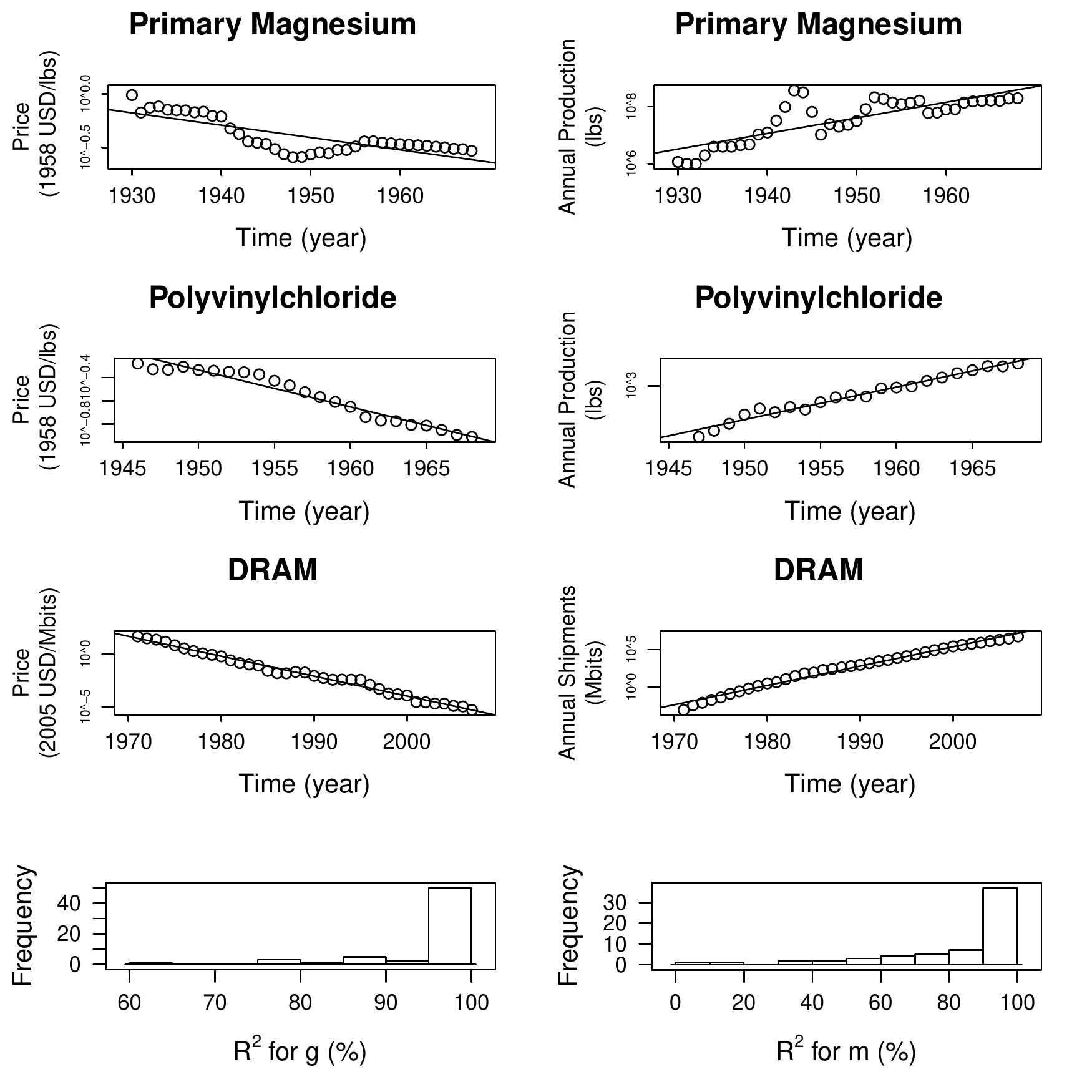}}
\caption{{\bf Three examples showing the logarithm of price as a function of time in the left column and the logarithm of production as a function of time in the right column, based on industry-wide data.}  We have chosen these examples to be representative:  The top row contains an example with one of the worst fits, the second row an example with an intermediate goodness of fit, and the third row one of the best examples. The fourth row of the figure shows histograms of $R^{2}$ values for fitting $g$ and $m$ for the 62 datasets.}
\label{r2gmw0ex}
\end{figure}

Although Moore performs slightly worse than Wright, given the clear difference in their economic interpretation, it is surprising that their  performance is so similar.  A simple explanation for Wright's law in terms of Moore's law was originally put forward by Sahal \cite{Sahal79}.  
He noted that if cumulative production grows exponentially\footnote{
Note that if production grows exponentially, cumulative production also grows exponentially with the same exponent.},
i.e.
\begin{equation}
x_t = A \exp(gt),
\label{production}
\end{equation}
then eliminating $t$ between Eqs.~(\ref{Moore}) and (\ref{production}) results in Wright's law, Eq.~(\ref{Wright}), with $w = m/g$.  Indeed, when we look at production vs. time we find that in almost every case the cumulative production increases roughly exponentially with time.  This is illustrated in Fig.~\ref{r2gmw0ex}, where we show three representative examples for production and cost plotted as a function of time. Fig.~\ref{r2gmw0ex} also shows histograms of $R^{2}$ values for fitting $g$ and $m$ for the 62 datasets. The agreement with exponential behavior ranges from very good to rather poor, but of course these are short time series and some of them are very noisy.

\begin{figure}
\vspace{2em}
\centering
\hspace{-2em}
\includegraphics[scale=0.4]{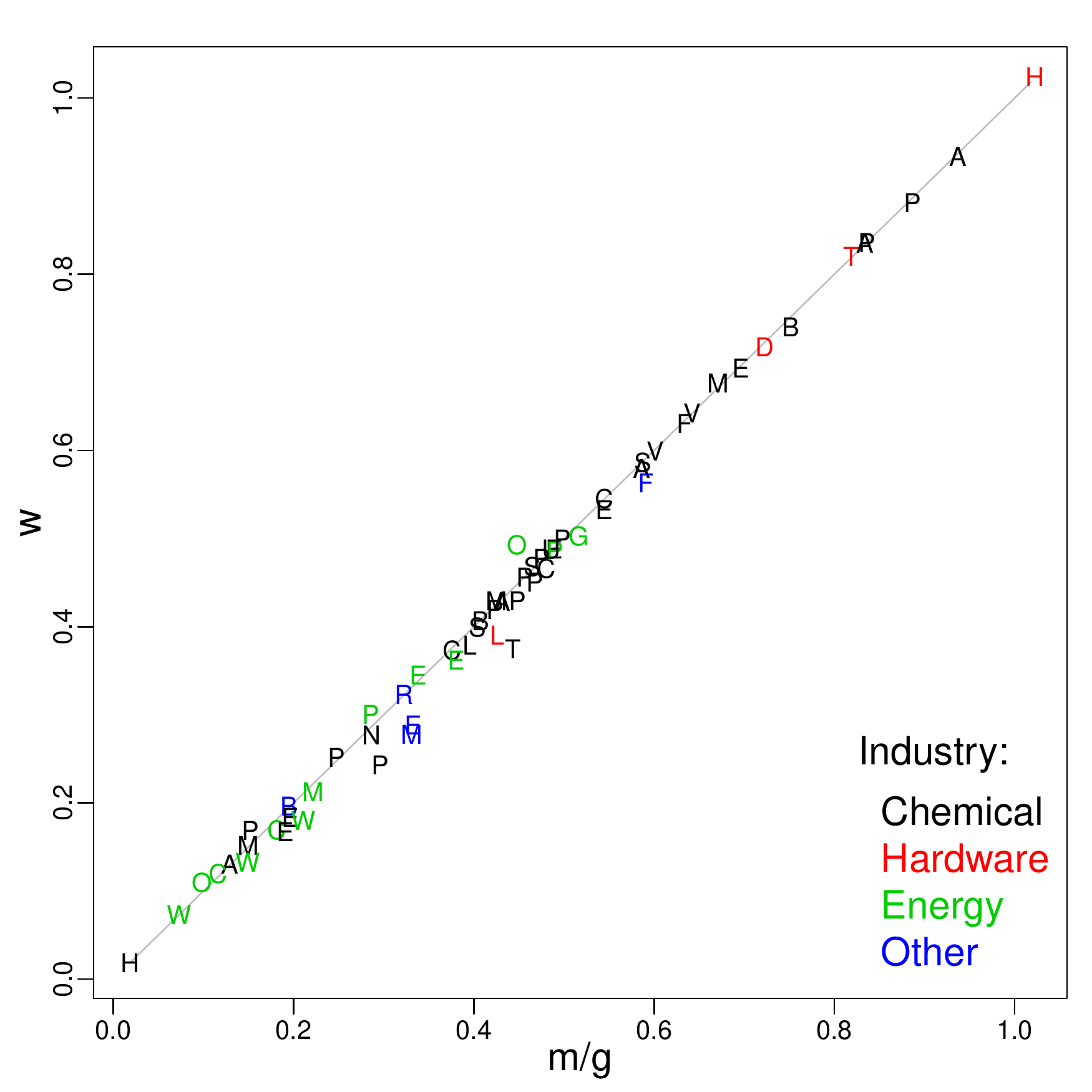} 
\caption{{\bf An illustration that the combination of exponentially increasing production and exponentially decreasing cost are equivalent to Wright's law.}  The value of the Wright parameter $w$ is plotted against the prediction $m/g$ based on the Sahal formula, where $m$ is the exponent of cost reduction and $g$ the exponent of the increase in cumulative production. }
\label{SahalFig}
\end{figure}

We test this in Fig.~\ref{SahalFig} by plotting the measured value of $w_d$  against  the derived value $\hat{w}_d = m/g$ for each data set $d$.  
The values cluster tightly along the identity line, indicating that Sahal's conjecture is correct.

\begin{figure}
\vspace{2em}
\centering
\hspace{-2em}
\includegraphics[scale=0.4]{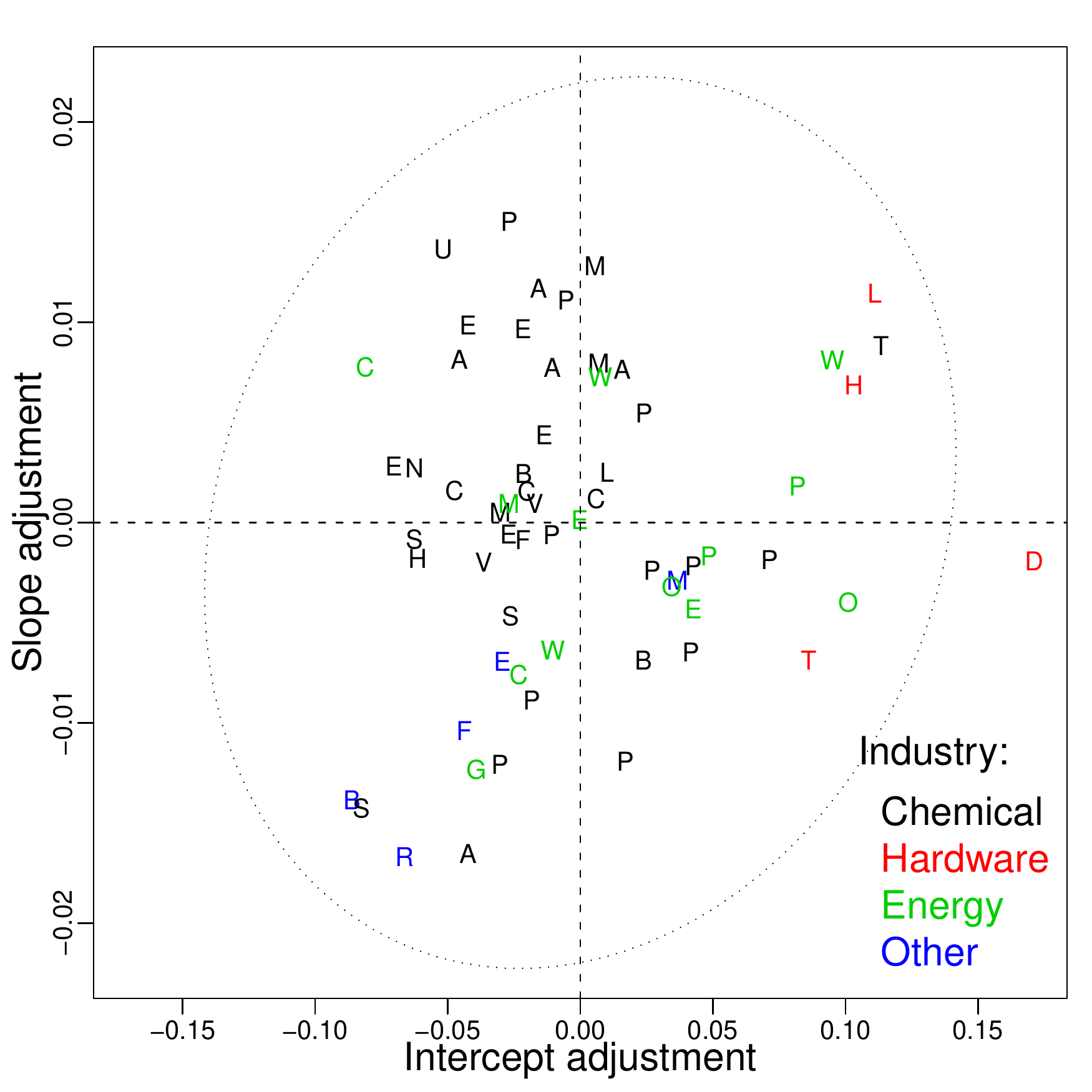} 
\caption{{\bf An illustration of how individual datasets deviate from the pooled data.}  The data-specific contribution to the slope, $b_d$, is plotted against the data specific contribution to the intercept, $a_d$, and compared to the ellipse of two standard deviation errors.  The best forecasts are obtained for those found in the lower left quadrant, such as Beer, Sodium, RefinedCaneSugar, and Aluminum.}
\label{groupsAdjustments}
\end{figure}

The differences in the data sets can be visualized by plotting $a_d$ and $b_d$ as shown in Fig.~\ref{groupsAdjustments}. All but one data set is inside the 95\% confidence ellipsoid, indicating that the estimated distribution of $(a_d, b_d)$ is consistent with the bivariate normal assumption.  The intercepts vary in a range roughly $-0.10 < a_d < 0.17$ and the slopes $-0.018 < b_d < 0.015$.  Thus the variation in the corresponding logarithmic forecasting error for the different datasets is comparable to the average error for all datasets (Fig.~\ref{groupsAdjustments}) and about an order of magnitude larger than the difference between the hypothesized laws (Fig.~\ref{fits}).

\begin{figure}
\vspace{2em}
\centering
\hspace{-2em}
\includegraphics[scale=0.4]{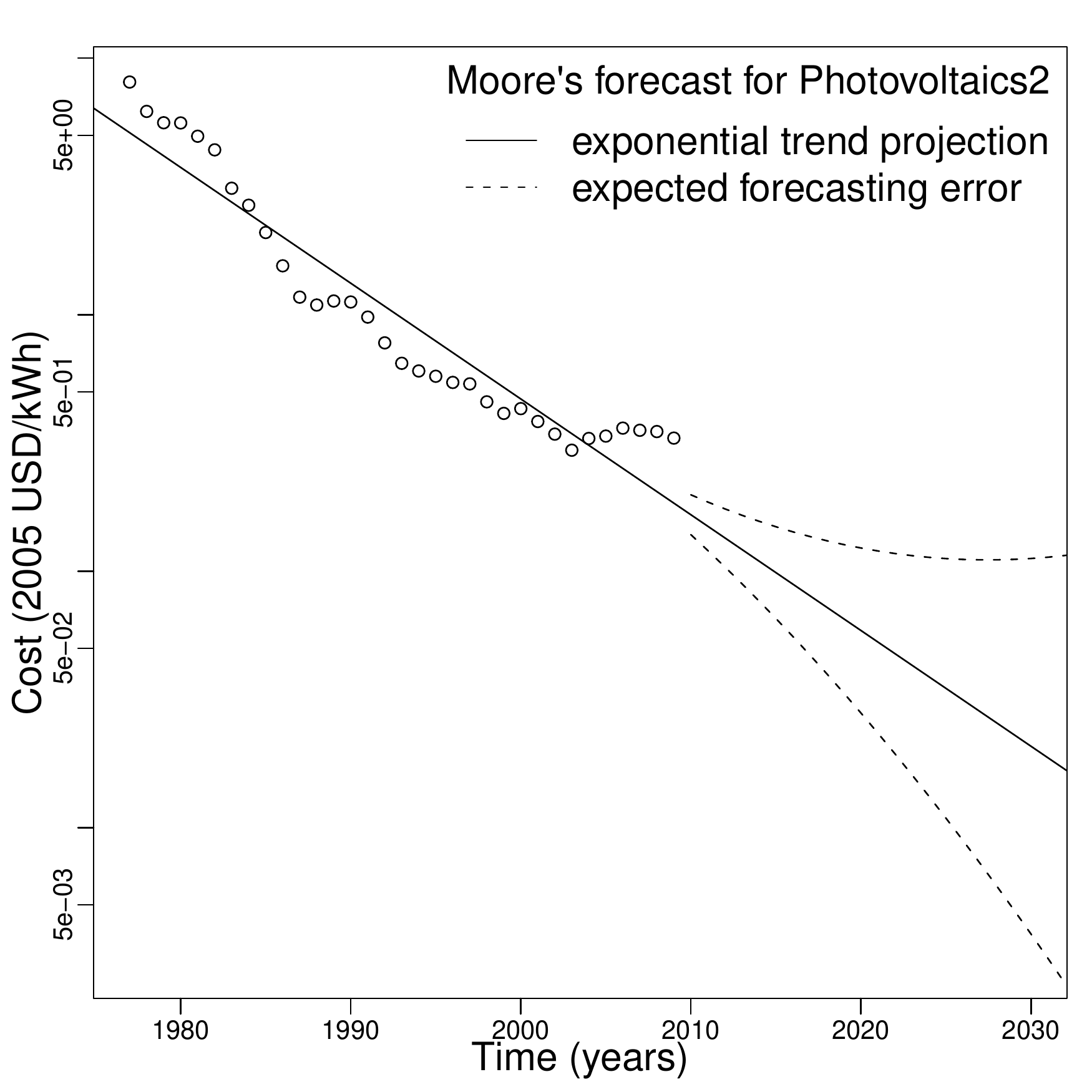} 
\caption{{\bf A projection of future PV electricity costs from the Photovoltaics2 historical data set (1977 -- 2009) using Moore's exponential functional form.}  The solid line is the expected forecast and the dashed line is the expected error.}
\label{pv}
\end{figure}

To illustrate the practical usefulness of our approach we make a forecast of the cost of electricity for residential scale photovoltaic solar systems (PV). Fig.~\ref{pv} shows the best forecast (solid line) as well as the expected error (dashed lines)\footnote{
Note that these are not confidence limits, but rather projected absolute log deviations from the best forecast, calculated from Eq. (\ref{errorModel}) using $\alpha_{\mbox{\footnotesize Moore}}$,  $\beta_{\mbox{\footnotesize Moore}}$,  $a_{Photovoltaics2}$, and $b_{Photovoltaics2}$.  The sharp drop in the one year forecast relative to the last observed data point comes from the fact that  forecasts are based on the average trend line, and because this data series is particularly long.  PV costs have risen recently due to increased material costs and other effects, but many industry experts expect this to be a short-lived aberration from the long-term cost trend.   See footnote 4 and Section 5 of the Supporting Information.}.
The expected cost in 2020 is 6 cents/kWh, with a range (3, 12), and in 2030 it is 2 cents, with a range (0.4, 11)\footnote{
This does not include the additional cost of energy storage technologies. Note also that this is for residential scale PV.  Industrial scale PV is typically about two-thirds the cost of electricity from residential scale systems.}. 
The current cost of the cheapest alternative, coal-fired electricity, is roughly 5 cents/kWh\footnote{
This is the wholesale cost at the plant (busbar), which may be most directly comparable to industrial scale PV.}.
In contrast to PV, coal-fired electricity is not expected to decrease in cost, and will likely increase if there are future penalties for CO$_{2}$ emissions \cite{McNerney10}.

%Wide-spread adoption of PV would require energy storage which would increase cost, but a similar method to that outlined here could be used to forecast the cost evolution of various energy storage technologies. While the near-term forecast in Fig.~\ref{pv} is off, our model predicts that we should eventually return to the trendline shown with the expected error we quantify based on past data. (Incidentally, experts in the PV industry also consider the recent cost hump as a short term artifact, and it is interesting that our very different, statistical approach leads us to the same conclusion.)

%An alternative perspective would be obtained by forecasting with Wright's law, which would suggest a level of effort or investment needed to reach a given cost. Using our method, the expectation based on both Moore and Wright can be used to inform decisions about the cost of a climate policy. There is currently substantial disagreement about the representation of technologies in these models, leading to highly variable estimates of policy costs. 

%An alternative perspective is obtained by forecasting with Wright's law.  It is useful to state the cumulative production in terms of the total expenditure $\sum_{i=1}^t y_t q_t$, i.e. the total expenditure for production.    

The key postulate that we have made here is that the processes generating the costs of technologies through time are generic except for technology-specific differences in parameters.   This hypothesis is powerful in allowing us to to view any given technology as being drawn from an ensemble.  This means that we can pool data from different technologies to make better forecasts, and most importantly, make error estimates. This is particularly useful in studying technology trends, where available data is limited.
%To illustrate why this is useful, we return again to the example of energy and climate policy. A policy maker relying on historical data for the roughly ten energy supply technology choices available is limited in the data analysis he/she can do on energy technologies alone. If one can draw on a greater pool of all technologies, forecasts can be much improved. 
Of course we must add the usual caveats about making forecasts -- as Neils Bohr reputedly said, prediction is very difficult, especially of the future.  
%But error estimates make our forecasts much more useful, and give an idea of how good we can expect them to be.

Our analysis reveals that decreasing costs and increasing production appear to be closely related, and that the hypotheses of Wright and Moore are more similar than they might appear.  We should stress, though, that they are not the same.  For example, consider a scenario in which the exponential rate of growth of PV production suddenly increased, which would decrease the current production doubling time of roughly 3 years. In this case, Wright predicts that the rate at which costs fall would increase, whereas Moore predicts that it would be unaffected. Distinguishing between the two hypotheses requires a sufficient number of examples where production does not increase exponentially, which our current database does not contain. The historical data shows a strong tendency, across different types of technologies, toward constant exponential growth rates. Recent work has, however, demonstrated super-exponential improvement for information technologies \cite{Nagy10b} over long time spans, suggesting that Moore's law is only a reasonable approximation over short spans of time. This evidence from information technology, and the results presented here, suggest that Moore may perform significantly worse than Wright over longer time horizons.

%\subsection*{Subsection 1}

%\subsection*{Subsection 2}

%\section*{Discussion}

% Do NOT remove this, even if you are not including acknowledgments
\section*{Acknowledgments}
This work was funded under National Science Foundation grant NSF-0738187.  All the conclusions and opinions are those of the authors and do not necessarily reflect those of the NSF. Partial funding for this work was provided by The Boeing Company (Contact: Scott H. Mathews, TF). We thank all contributors to the performance curve database (pcdb.santafe.edu).

%\section*{References}
% The bibtex filename

\pagebreak
\section*{Figure Legends}
\begin{figure}[!ht]
\begin{center}
\includegraphics[width=4in]{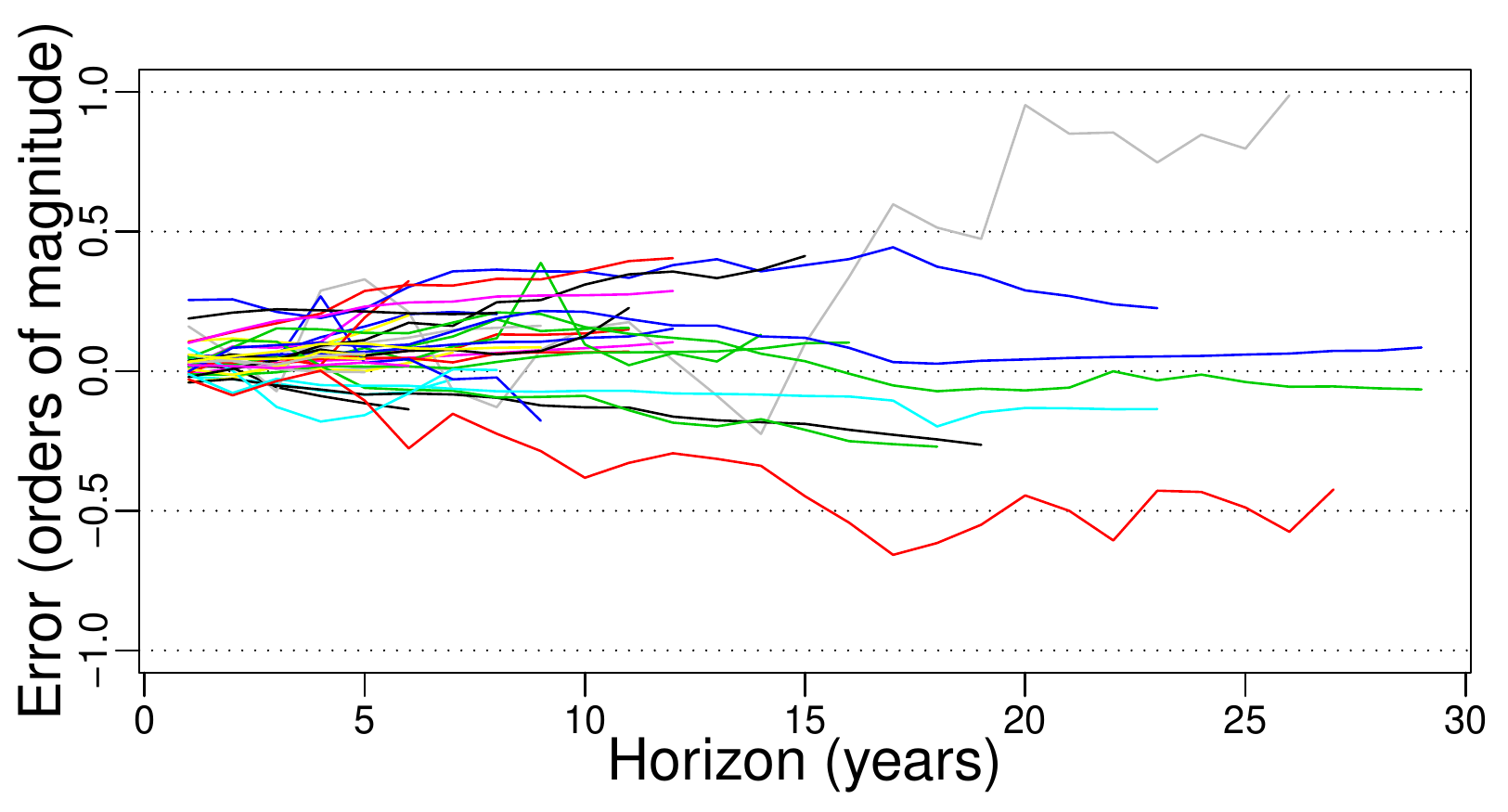}
\end{center}
\caption{
{\bf An illustration of the growth of errors with time using the Wright model.}  The mean value of the logarithmic hindcasting error for each dataset is plotted against the hindcasting horizon $j - i$, in years.  An error of $10^{0.5} \approx 3$, for example, indicates that the predicted value is three times as big as the actual value. The longest data-sets are:  PrimaryAluminum (green), PrimaryMagnesium (dark blue), DRAM (grey), and Transistor (red).
}
\label{Figure_label}
\end{figure}

%\section*{Tables}
%\begin{table}[!ht]
%\caption{
%\bf{Table title}}
%\begin{tabular}{|c|c|c|}
%table information
%\end{tabular}
%\begin{flushleft}Table caption
%\end{flushleft}
%\label{tab:label}
% \end{table}

\end{document}